\documentclass[aps,showpacs,floatfix]{revtex4}
\usepackage{amssymb,amsmath}
\usepackage[dvips]{graphicx,color}

\def\openone{\leavevmode\hbox{\small1\kern-3.8pt\normalsize1}}

\def\bea{\begin{eqnarray}}
\def\eea{\end{eqnarray}}
\def\beq{\begin{equation}}
\def\eeq{\end{equation}}

\begin{document}

\title{On covariant nonlocal chiral quark models with separable interactions}
\author{
D. G\'omez Dumm$^{a,b}$,
A.G. Grunfeld$^c$ and
N.N. Scoccola$^{b,c,d}$}
\affiliation{
$^a$ IFLP $-$ Dpto.\ de F\'{\i}sica,
     Universidad Nacional de La Plata,
     C.C.\ 67, 1900 La Plata, Argentina \\
$^b$ CONICET, Rivadavia 1917, 1033 Buenos Aires, Argentina\\
$^c$ Physics Department, Comisi\'on Nacional de Energ\'{\i}a At\'omica,
 Av.\ Libertador 8250, 1429 Buenos Aires, Argentina \\
$^d$ Universidad Favaloro, Sol{\'\i}s 453, 1078 Buenos Aires,
Argentina}

\begin{abstract}
We present a comparative analysis of chiral quark models which include
nonlocal covariant four-fermion couplings. We consider two alternative
ways of introducing the nonlocality, as well as various shapes for the
momentum-dependent form factors governing the effective interactions. In
all cases we study the behavior of model parameters and analyze numerical
results for constituent quark masses and quark propagator poles.
Advantages of these covariant nonlocal schemes over instantaneous nonlocal
schemes and the standard NJL model are pointed out.
\end{abstract}


\maketitle

\renewcommand{\thefootnote}{\arabic{footnote}}
\setcounter{footnote}{0}

\section{Introduction}

The description of strong interactions in the nonperturbative
regime is still one of the most important open problems in
particle physics. It is believed that the strong interaction
Lagrangian supports an SU(2) chiral symmetry which is dynamically
broken at low energies, and pions play the r\^ole of the
corresponding Goldstone bosons. A simple scheme including these
properties is the well known Nambu$-$Jona-Lasinio (NJL)
model~\cite{Nambu:1961tp}, proposed more than four decades ago.
The NJL model has been widely used as an schematic effective
theory for QCD~\cite{Vogl:1991qt,Klevansky:1992qe,Hatsuda:1994pi},
allowing e.g.\ the description of light mesons as
fermion-antifermion composite states.

In the NJL model quarks interact through a local, chiral invariant
four fermion coupling. Due to the local nature of this
interaction, the corresponding Schwinger-Dyson and Bethe-Salpeter
equations become relatively simplified. However, the main
drawbacks of the model are direct consequences of this locality:
loop integrals are divergent (and therefore have to be regulated
somehow), and the model is nonconfining. The absence of
confinement is basically related to the fact that the dynamically
generated constituent quark masses are momentum independent. This
imposes a severe restriction on the range of applicability of the
model, since a $\bar q q$ continuum opens up at energies of twice
the constituent mass. Since the NJL model is nonrenormalizable, in
practice it is necessary to apply some form of ultraviolet
regularization, with the introduction of a finite cutoff
parameter. The adopted regularization scheme has to be regarded as
a part of the model itself. A variety of schemes have been used in
the literature, such as sharp three- or four-momentum cutoffs, and
proper time or Pauli-Villars regulators. Although the model does
contain regularization-independent
information~\cite{Bijnens:1993ap}, and results obtained within
various regularization schemes are found to be qualitatively
similar~\cite{Meissner:1990se}, the choice of any particular
scheme lacks a strong physical motivation. Moreover, it can be seen
that one needs further regularization prescriptions in order to
carry out calculations beyond the leading order in the $1/N_c$
expansion (a new cutoff is required for meson
loops~\cite{Dmitrasinovic:1995cb}). On the other hand, the
introduction of a finite cutoff is somewhat problematic in the
anomalous sector: if low energy theorems for anomalous processes
such as $\pi^0 \rightarrow 2 \gamma$ are assumed to be valid, then
the corresponding quark loop integrals should include a complete
set of quark states. In the NJL framework, this can be achieved
either by leaving the anomalous diagrams ad-hoc
unregulated~\cite{Schuren:1993aj} or by including additional terms
in the Lagrangian in order to recover the anomalous Ward
identities~\cite{Bijnens:1993ap}.

As a way to improve upon the NJL model, extensions which include nonlocal
interactions have been proposed (see Ref.~\cite{Rip97} and references
therein). In fact, nonlocality arises naturally in the context of several
well established approaches to low energy quark dynamics, as e.g.\ the
instanton liquid model~\cite{SS98} and the Schwinger-Dyson resummation
techniques~\cite{Roberts:1994dr}. Lattice QCD
calculations~\cite{Parappilly:2005ei} also indicate that quark
interactions should act over a certain range in the momentum space.
Moreover, it has been argued that nonlocal extensions of the NJL model do
not show some of the above mentioned inconveniences of the local theory.
Indeed, nonlocal interactions regularize the model in such a way that
anomalies are preserved~\cite{RuizArriola:1998zi} and charges are properly
quantized, the effective interaction is finite to all orders in the loop
expansion and therefore there is not need to introduce extra
cutoffs~\cite{Blaschke:1995gr}, soft regulators such as Gaussian functions
lead to small next-to-leading order corrections~\cite{Rip00}, etc. In
addition, it has been shown~\cite{Bowler:1994ir,Plant:1997jr} that a
proper choice of the nonlocal form factor and the model parameters can
lead to some form of quark confinement, in the sense that the effective
quark propagator has no poles at real energies~\cite{Sti87}.

In order to bring the problem into a tractable form, most of the
calculations reported in the literature deal with nonlocal
interactions which are separable in momentum space. In fact,
basically two alternative schemes to introduce nonlocality in a
separable way have been considered. One of
them~\cite{Buballa:1992sz,Bowler:1994ir,Plant:1997jr} is inspired
on the instanton liquid picture of QCD, which indeed gives rise to
an effective separable nonlocal four quark vertex. The other
one~\cite{Ito:1991sz,Schmidt:1994di} might be understood as a
separable approximation to an effective one-gluon exchange
interaction. Considerable work has been done using one scheme or
the other. This includes studies of the
mesonic~\cite{Bowler:1994ir,Plant:1997jr,Scarpettini:2003fj,Noguera:2005ej}
and the baryonic~\cite{Golli:1998rf,Rezaeian:2004nf} sectors, as
well as the behavior of quark matter at finite temperature and
densities~\cite{General:2000zx}. The aim of the present work is to
present a detailed comparison of both approaches and their
relation with the NJL model, analyzing the respective input
parameter ranges for different form factors and discussing the
validity of low energy theorems.

The article is organized as follows. In Sec.\ II we introduce a
common framework to deal with the above mentioned approaches to
nonlocal separable interactions. Analytical expressions for mean
field quantities, meson masses and decay constants are given for
both schemes. In Sec.\ III, the validity of various chiral
relations is explicitly shown. In Sec.\ IV we quote some numerical
results for definite nonlocal form factors. We consider here the
widely used Gaussian form factor, as well as a family of
Lorentzian functions which allow us to interpolate between soft
form factors and sharp NJL-like cutoffs. Finally, our main results
and conclusions are outlined in Sec.\ V.

\section{Covariant nonlocal models with separable interactions}

\subsection{Effective action}

Let us begin by stating the Euclidean action for the nonlocal chiral quark
model in the case of two light flavors,
\begin{equation}
S_E = \int d^4 x \ \left\{ \bar \psi (x) \left(- i \rlap/\partial
+ m_c \right) \psi (x) -
\frac{G}{2} j_a(x) j_a(x) \right\} \ .
\label{action}
\end{equation}
Here $m_c$ is the current quark mass, which is assumed to be equal
for $u$ and $d$ quarks. As mentioned in the Introduction,
separable nonlocal interactions might be introduced in two
alternative ways. In what follows, we will call them ``Scheme~I''
and ``Scheme~II''. These schemes are characterized by the form of
the nonlocal currents $j_a(x)$ in Eq.~(\ref{action}). In the case
of Scheme~I~\cite{Buballa:1992sz,Bowler:1994ir,Plant:1997jr}, the
effective interactions are based in an instanton liquid picture of
QCD. The currents can be written as
\begin{eqnarray}
j_a (x) &=& \int d^4 y\ d^4 z \ r(y-x) \ r(x-z) \ \bar \psi(y) \
\Gamma_a \ \psi(z) \ ,
\label{uno}
\end{eqnarray}
where we have defined $\Gamma_a = ( \openone, i \gamma_5 \vec \tau )$,
$\tau_i$ being the Pauli matrices acting on flavor space.

On the other hand, Scheme II~\cite{Ito:1991sz,Schmidt:1994di} arises from a
separable form of the effective one-gluon exchange picture. In this case
the nonlocal currents $j_a(x)$ are given by
\begin{eqnarray}
j_a (x) &=& \int d^4 z \  g(z) \
\bar \psi(x+\frac{z}{2}) \ \Gamma_a \ \psi(x-\frac{z}{2}) \ .
\label{cuOGE}
\end{eqnarray}
The functions $r(x-y)$ and $g(z)$ in Eqs.~(\ref{uno}) and (\ref{cuOGE}),
respectively, are nonlocal form factors characterizing the corresponding
interactions. It is convenient to translate them into momentum space.
Since Lorentz invariance implies that they can only be functions of $p^2$,
we will use for the Fourier transforms of these regulators the forms
$r(p^2)$ and $g(p^2)$ from now on. A schematic representation of the four
quark interactions in both schemes is given in Fig.~1.

\begin{figure}[htb]
\begin{center}
\centerline{ \includegraphics[height=7truecm]{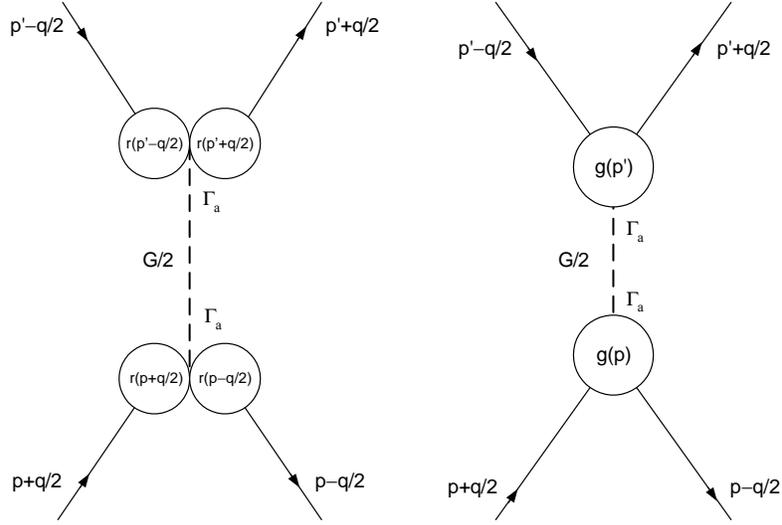} } \caption{Schematic
representation of the four fermion interaction in Scheme I (left) and Scheme
II (right).} \label{inter}
\end{center}
\end{figure}

In order to deal with meson degrees of freedom, one can perform a
standard bosonization of the theory. This is done by considering
the corresponding partition function $Z = \int {\cal D} \bar \psi
\,{\cal D} \psi \, \exp[- S_E]$, and introducing auxiliary fields
$S_a(x) = ( \sigma(x), \vec \pi(x) )$, where $\sigma(x)$ and $\vec
\pi(x)$ are scalar and pseudoscalar mesons, respectively.
Integrating out the quark fields we get
\begin{eqnarray}
Z &=& \int {\cal D} \sigma\, {\cal D} \vec \pi \
\exp[- S_E^{\rm bos}] \ ,
\end{eqnarray}
where
\begin{eqnarray}
S_E^{\rm bos} = - \ln\, \det A + \frac{1}{2 G} \int \frac{d^4 p}{(2 \pi)^4}
\ S_a(p) \ S_a(-p) \ .
\end{eqnarray}
The operator $A$ reads, in momentum space,
\begin{eqnarray}
A(p,p') &=& (\,-\rlap/p + m_c)\,(2\pi)^4 \,\delta^{(4)}(p-p')
 + h(p,p')\ \Gamma_a \ S_a(p-p') \ ,
\end{eqnarray}
where
\begin{eqnarray}
h(p,p') \ = \ \left\{\begin{array}{lcl}
r(p^2)\ r(p'^2) & \qquad \qquad & {\rm (Scheme\ I)} \\
\\
g\Big[\,\Big(\frac{p+p'}{2}\Big)^2\,\Big] & &
{\rm (Scheme\ II)} \ \ . \\
\end{array}
\right.
\end{eqnarray}

At this stage we assume that the $\sigma$ field has a nontrivial
translational invariant mean field value $\bar \sigma$, while the mean
field values of the pseudoscalar fields $\pi_i$ are zero. Thus we write
\begin{eqnarray}
\sigma(x) &=& \bar \sigma + \delta \sigma(x) \\
\vec \pi(x) &=& \delta \vec \pi (x)
\end{eqnarray}
Replacing in the bosonized effective action, and expanding in powers of
the meson fluctuations, we get
\begin{eqnarray}
S_E^{\rm bos} \ = \ S_E^{\rm MFA} + S_E^{\rm quad} + \ ...
\end{eqnarray}
Here the mean field action per unit volume reads
\begin{eqnarray}
\frac{S_E^{\rm MFA}} {V^{(4)}} &=& - 4 N_c  \int \frac{d^4 p}{(2\pi)^4} \
\ln \left[ p^2 + \Sigma^2(p) \right] + \frac{\bar \sigma^2}{2 G} \ ,
\end{eqnarray}
where $\Sigma (p) = m_c + h(p,p)\ \bar \sigma$. The quadratic terms are
given by
\begin{equation}
S_E^{\rm quad} = \frac{1}{2} \int \frac{d^4p}{(2\pi)^4} \left[
G^+(p^2) \ \delta\sigma(p) \ \delta\sigma(-p) +
G^-(p^2) \ \delta\vec\pi(p)    \cdot  \delta\vec\pi(-p) \right] \ ,
\end{equation}
where
\begin{eqnarray}
G^{\pm}(p^2) = \frac{1}{G} - \, 8 \,N_c \int \frac{d^4 q}{(2 \pi)^4}\
 h^2(q^+,q^-) \frac{  \left[ (q^+ \cdot q^-) \mp \Sigma(q^+)
 \Sigma(q^-)\right]}{\left[ (q^+)^2 + \Sigma^2(q^+) \right]
 \left[ (q^-)^2 + \Sigma^2(q^-)\right]}
\label{vamopincha}
\end{eqnarray}
with $q^\pm = q \pm p/2\,$.

\subsection{Mean field approximation and chiral condensates}

In order to find the mean field values $\bar \sigma$, one has to minimize
the action $S_E^{\rm MFA}$. A straightforward exercise leads to the
gap equation
\begin{eqnarray}
\bar \sigma - 8 N_c \ G \int \frac{d^4 p}{(2 \pi)^4}\  h(p,p) \ \frac{
\Sigma(p) } {p^2 + \Sigma^2(p)} &=& 0 \ .
\label{gapeq}
\end{eqnarray}

Now the chiral condensates are given by the vacuum expectation values
$\langle\bar q q\rangle = \langle\bar u u\rangle = \langle\bar d
d\rangle$. They can be easily obtained by performing the variation of
$Z^{\rm MFA} = \exp[ -S_E^{MFA} ]$ with respect to the corresponding
current quark masses. We obtain
\begin{eqnarray}
\langle\, \bar q\, q\, \rangle &=& - \, 4 N_c \int \frac{d^4 p}{(2 \pi)^4}\
\frac{\Sigma(p)} {p^2 + \Sigma^2(p)} \ \ .
\label{cond}
\end{eqnarray}

\subsection{Meson masses and quark-meson coupling constants}

In what follows we consider in particular the case of pseudoscalar mesons.
The expressions for scalar mesons are completely equivalent, just
replacing the upper index ``$-$'' by ``$+$'' where it corresponds.

The pion mass can be obtained by solving the equation
\bea
G^-(-m_\pi^2) = 0 \ .
\label{pieq}
\eea
Now, one has to perform a wave function renormalization of the
pseudoscalar fields. The renormalized pion field $\vec {\tilde \pi}(p) =
\vec \pi (p) / Z_\pi^{1/2}$ is defined by fixing the residue of
$G^-(p^2)$ at the pion pole, i.e.\ by demanding that in the vicinity of
the pole the corresponding contribution to the quadratic lagrangian is
given by
\begin{equation}
\left({\cal L}_E^{\rm quad}\right)_\pi  =
\quad \frac12 \left( p^2 + m^2_\pi \right) \ {\vec {\tilde \pi}}(p)
\cdot {\vec {\tilde \pi}}(-p) \ .
\end{equation}
In this way, one obtains
\begin{equation}
Z_\pi^{-1} \ = \ \frac{d G^-(p) }{dp^2}
\bigg|_{p^2=-m_\pi^2} \ .
\label{zpi}
\end{equation}
Finally the $\pi q\bar q$ coupling constant $G_{\pi q \bar q}$ is given by
the residue of the pion propagator at the pole, leading to
\begin{equation}
G_{\pi q \bar q} \ = \ Z^{1/2}_{\pi} \ .
\label{gpiqq}
\end{equation}

\subsection{Pion weak decay constant}

By definition the pion weak decay constant $f_\pi$ is given by the matrix
element of the axial current $A^a_\mu(x)$ between the vacuum and the
renormalized one-pion state at the pion pole:
\begin{equation}
\langle 0 | A^a_\mu(0) | \tilde \pi^b (p) \rangle = i \ \delta^{ab} \
p_\mu \ f_\pi \ .
\end{equation}

In order to obtain an explicitly expression for the axial current, we have
to ``gauge'' the effective action $S_E$ by introducing a set of axial
gauge fields ${\cal A}^a_\mu(x)$. For a local theory this ``gauging''
procedure is usually done by performing the replacement
\begin{equation}
\partial_\mu \rightarrow \partial _\mu + \frac{i}{2} \ \gamma_5
\ \vec \tau \cdot {\vec {\cal A}}_\mu(x) \ .
\end{equation}
In the present case ---owing to the nonlocality of the involved fields---
one has to perform an additional replacement, namely
\begin{eqnarray}
\gamma_0 \ \Gamma_a \ S_a(x) \ \rightarrow
\ \left\{ \begin{array}{lcl}
W_A\left( y , x \right) \ \gamma_0 \ \Gamma_a \ S_a(x) \ W_A\left(x, z
\right) & \qquad \qquad & {\rm (Scheme\ I)} \\
\rule{0cm}{.7cm} W_A\left( x + z/2 , x \right) \
\gamma_0 \ \Gamma_a \ S_a(x) \ W_A\left(x, x - z/2 \right)
& & {\rm (Scheme\ II)\ .}
\end{array}
\right.
\label{gauge}
\end{eqnarray}
Here $y$ and $z$ are the integration variables in the definitions of the
nonlocal currents (see Eqs.~(\ref{uno}) and (\ref{cuOGE})), and the
function $W_A(x,y)$ is defined by
\begin{equation}
W_A(x,y) \ = \ {\rm P}\; \exp \left[\frac{i}{2} \int_x^y ds_\mu
\ \gamma_5 \ \vec \tau  \cdot {\vec {\cal A}}_\mu(s) \right] \ ,
\end{equation}
where $s$ runs over an arbitrary path connecting $x$ with $y$.
This corresponds to the introduction of a parallel
transport~\cite{BKN91}.

Once the gauged effective action is built, it is easy to get the axial
current as the derivative of this action with respect to ${\cal
A}^a_\mu(x)$, evaluated at ${\vec{\cal A}}_\mu(x)=0$. Performing the
derivative of the resulting expressions with respect to the renormalized
meson fields, we can finally identify the corresponding meson weak decay
constants. After a rather lengthy calculation we obtain
\begin{equation}
f_\pi \ = \ Z_\pi^{1/2} \ \frac{F(-m_\pi^2) - F(0)}{m_\pi^2} \ ,
\label{fpieq}
\end{equation}
where $F(p^2)$ is given by
\begin{equation}
F(p^2) \ = \ m_c \; J(p^2) \ + \ \bar \sigma \; K(p^2) \ ,
\label{jk}
\end{equation}
with
\begin{eqnarray}
J(p^2) & = & 8 \,N_c \int \frac{d^4 q}{(2 \pi)^4}\
 h(q^+,q^-) \; \frac{  \left[ (q^+ \cdot q^-) + \Sigma(q^+)
 \Sigma(q^-)\right]}{\left[ (q^+)^2 + \Sigma^2(q^+) \right]
 \left[ (q^-)^2 + \Sigma^2(q^-)\right]} \nonumber \\
K(p^2) & = & 8 \,N_c \int \frac{d^4 q}{(2 \pi)^4}\
 h^2(q^+,q^-) \; \frac{  \left[ (q^+ \cdot q^-) + \Sigma(q^+)
 \Sigma(q^-)\right]}{\left[ (q^+)^2 + \Sigma^2(q^+) \right]
 \left[ (q^-)^2 + \Sigma^2(q^-)\right]}
 \label{integrals}
\end{eqnarray}
(here, as before, $q^\pm = q \pm p/2$).

Notice that in terms of the functions $J(p^2)$ and $K(p^2)$
Eqs.~(\ref{gapeq}) and (\ref{pieq}) can be written as
\begin{eqnarray}
F(0) & = & \bar\sigma \,/\, G  \nonumber \\
K(-m_\pi^2) & = & 1\,/\, G \ .
\end{eqnarray}
Replacing these expressions in Eq.~(\ref{fpieq}), we get the relation
\begin{equation}
m_\pi^2 \; f_\pi \ = \ m_c \; Z_\pi^{1/2}\; J(-m_\pi^2) \ .
\label{relation}
\end{equation}

\subsection{$\pi^0 \to \gamma\gamma$ decay}

Let us analyze in our context the anomalous decay of the $\pi^0$ meson
into two photons. In general, the amplitude for this process can be
written as
\begin{equation}
A(\pi^0\to\gamma\gamma) \ =
\ 4\pi\,\alpha \; g_{\pi\gamma\gamma}\ \epsilon_{\mu\nu\alpha\beta}
\ \varepsilon_1^{\ast\mu}\;\varepsilon_2^{\ast\nu} \;k_1^\alpha\;k_2^\beta\ ,
\end{equation}
where $\alpha$ is the fine structure constant and $k_i$, $\varepsilon_i$
stand for the momenta and polarizations of the outgoing photons
respectively. The partial decay width is then given by
\begin{equation}
\Gamma(\pi^0\to \gamma\gamma) = \frac{\pi}{4}\;\alpha^2\; m_{\pi^0}^3\;
g_{\pi\gamma\gamma}^2\ .
\end{equation}

The coefficients $g_{\pi\gamma\gamma}$ are given by quark loop integrals.
Besides the usual ``triangle'' diagram, given by a closed quark loop with
one pion and two photon vertices, in the present nonlocal schemes one has
a second contribution~\cite{Plant:1997jr} in which one of the quark-photon
vertices arises from the gauge fields entering the nonlocal currents [see
Eq.~(\ref{gauge})]. The sum of both contributions leads to
\begin{equation}
g_{\pi\gamma\gamma} \ = \ I(-\,m_\pi^2)\ Z_\pi^{1/2}\ ,
\end{equation}
where
\begin{eqnarray}
I(-\,m_\pi^2) & = & \frac{8}{3}\; N_c \; \int\ \frac{d^4q}{(2\pi)^4}\ \
\frac{h(q+k_1,q-k_2)}
{[q^2+\Sigma^2(q)] \; [(q+k_1)^2+\Sigma^2(q+k_1)]
\; [(q-k_2)^2+\Sigma^2(q-k_2)]}\nonumber \\
& & \times \left\{ \Sigma(q)\ +\ \frac{q^2}{2}\;
\left[\frac{[\Sigma(q-k_2) - \Sigma(q)]}{(k_2\cdot q)} \ - \
\frac{[\Sigma(q+k_1) - \Sigma(q)]}{(k_1\cdot q)} \right] \right\}\ .
\label{iloop}
\end{eqnarray}
Notice that for on-shell photons the above integral is a function of the
scalar product $(k_1\cdot k_2)$ only, and in Euclidean space one has
$(k_1 \cdot k_2) = -\,m_\pi^2/2$.

\section{Chiral relations}

We start by discussing the Goldberger-Treiman (GT) relation, which states
that in the chiral limit ($\,m_c\rightarrow  0\,$) one has
\bea
f_{\pi,0} \ G_{\pi q\bar q,0} \ = \ \bar \sigma_0
\label{gt}
\eea
(here, and in the following, subindices $0$ mean that quantities are
evaluated in the chiral limit). In our case, from Eqs.~(\ref{fpieq}) and
(\ref{jk}) one has
\begin{equation}
f_{\pi,0} \ = \ Z_{\pi,0}^{1/2} \;
\bigg[\frac{F_0(-p^2)-F_0(0)}{p^2}\bigg]_{p^2=0} \ = \ -
\,\bar\sigma_0\; Z_{\pi,0}^{1/2} \;K'_0(0) \ ,
\label{gt1}
\end{equation}
where prime denotes derivation with respect to $p^2$. Now, an explicit
calculation shows that $f_{\pi,0}$ can be written in terms of the
momentum-dependent constituent masses $\Sigma(q)$ as
\begin{equation}
f^2_{\pi,0} = 2 N_c \int \frac{d^4 q}{(2 \pi)^4} \ \
 \frac{\; 2\,\Sigma^2_0(q) \, - \, q^2\,\Sigma_0(q)\, \Sigma'_0(q)}
  { [\, q^2 + \Sigma^2_0(q) \,]^2 } \ \ . \label{fpich}
\end{equation}
Then, taking into account Eqs.~(\ref{zpi}) and (\ref{gpiqq}), and noting
that $G^-_0(p^2) = 1/G - K_0(p^2)$, one gets
\begin{equation}
K'_0(0) \ = \ -\,Z_{\pi,0}^{-1} \ = \ -\,G_{\pi q\bar q,0}^{-2} \ .
\label{gt2}
\end{equation}
The GT relation (\ref{gt}) follows immediately from (\ref{gt1}) and
(\ref{gt2}).

Next, let us consider the Gell-Mann-Oakes-Renner (GOR) relation, which can
be obtained using Eq.~(\ref{relation}). By performing a chiral expansion,
and keeping only the lowest nonzero order at both sides of this equation,
one has
\begin{equation}
m_\pi^2 \; f_{\pi,0} \ = \ m_c \; Z_{\pi,0}^{1/2}\; J_0(0) \ .
\label{relquir}
\end{equation}
In addition, from Eq.~(\ref{cond}) it is easy to see that
\begin{equation}
\langle \,\bar u\, u + \bar d\, d \,\rangle_0 \ = \ 2\, \langle \,\bar q\,
q\,\rangle_0 \ = \ - \, \bar\sigma_0 \, J_0(0) \ ,
\end{equation}
thus using the GT relation one arrives to
\begin{equation}
m_c \ \langle\, \bar u\,  u + \bar d\,  d \,\rangle_0 \ = - \ f_{\pi,0}^2 \
m_\pi^2\ ,
\label{GOR}
\end{equation}
which is the form taken by the GOR relation in the isospin limit.

Finally, we study the anomalous effective coupling $g_{\pi\gamma\gamma}$
in the chiral limit. Considering Eq.~(\ref{iloop}), one can perform an
expansion in powers of the momenta $k_1$ and $k_2$ and take the limit
$m_\pi^2\to 0$. This leads to
\begin{equation}
I_0(0) \ = \ \frac{N_c}{12\pi^2\,\bar\sigma_0} \ .
\end{equation}
Now, for $N_c=3$,  using the GT relation one gets
\begin{equation}
g_{\pi\gamma\gamma,0}\ =\ \frac{1}{4\pi^2 f_{\pi,0}}\ ,
\label{FHrel}
\end{equation}
which is the expected result according to low energy theorems and Chiral
Perturbation Theory.

We stress that the derivation of the above chiral relations holds for both
schemes I and II.

\section{Comparative analysis for definite form factors}

In what follows we consider two types of form factors which have been
frequently used in the literature. One is the Gaussian form factor,
\begin{equation}
g_{\rm G}(p^2) = \Big[ r_{\rm G}(p^2) \Big]^2 = \exp( - p^2/\Lambda^2 ) \ .
\end{equation}
The other is the Lorenztian form factor
\begin{equation}
g_{\rm Ln}(p^2) = \Big[ r_{\rm Ln}(p^2) \Big]^2 =
\frac{1}{1 + \left(p^2/\Lambda^2\right)^n} \ ,
\end{equation}
where $n \ge 2$ is an integer that allows us to interpolate between
soft form factors (such as the Gaussian one) and sharp form factors as
e.g.\ a step function $g_{\rm Step}(p^2) = \Theta(\Lambda^2/p^2 - 1)$.

With the choice $g(p^2) = [ r(p^2) ]^2$, both schemes I and II lead to the
same expressions for the mean field quantities. Moreover, in the chiral
limit, the expressions for the pion decay constant converge in both cases to
the result in Eq.~(\ref{fpich}). This simplifies the comparison of the
present type of nonlocal models with other frameworks such as instantaneous
nonlocal schemes and the usual NJL model. In the following subsection we will
present a comparative analysis, considering in particular the chiral limit.
Results corresponding to the case of finite quark masses will be discussed in
subsection B.

\subsection{Chiral limit case}

In this limit the pion is massless. Moreover, a simple dimensional analysis
shows that, for the kind of models under study, the chiral condensate can be
expressed as
\begin{equation}
\left[ -\, \langle\, \bar q\,  q \rangle_0 \, \right]^{1/3} \ = \ f_{\pi,0}
\ {\cal F}(G \,\Lambda^2) \ .
\end{equation}
Thus, for a given value of $f_{\pi,0}$ the chiral condensate only depends on
a single dimensionless parameter, $G\Lambda^2$. This allows to choose a
typical value for the pion decay constant in the chiral limit, say $f_{\pi,0}
= 90$ MeV, and study the behavior of the chiral condensate as a function of
$G\Lambda^2$ for various models. We have performed this analysis, considering
the range of values for which the different approaches are compatible with
the empirical bounds $-\langle\bar q q \rangle^{1/3}\, \simeq \, 200$ to
$260$~MeV~\cite{Dosch:1997wb,Giusti:1998wy}. Our results are displayed in
Fig.~2. In the upper panel we show the curves corresponding to covariant
models, with different form factors. For comparison, in the lower panel we
show the results obtained for the case of instantaneous nonlocal
models~\cite{Schmidt:1994di,Grigorian:2006qe}, and for the usual NJL model.

\begin{figure}[hbt]
\begin{center}
\centerline{ \includegraphics[height=15truecm]{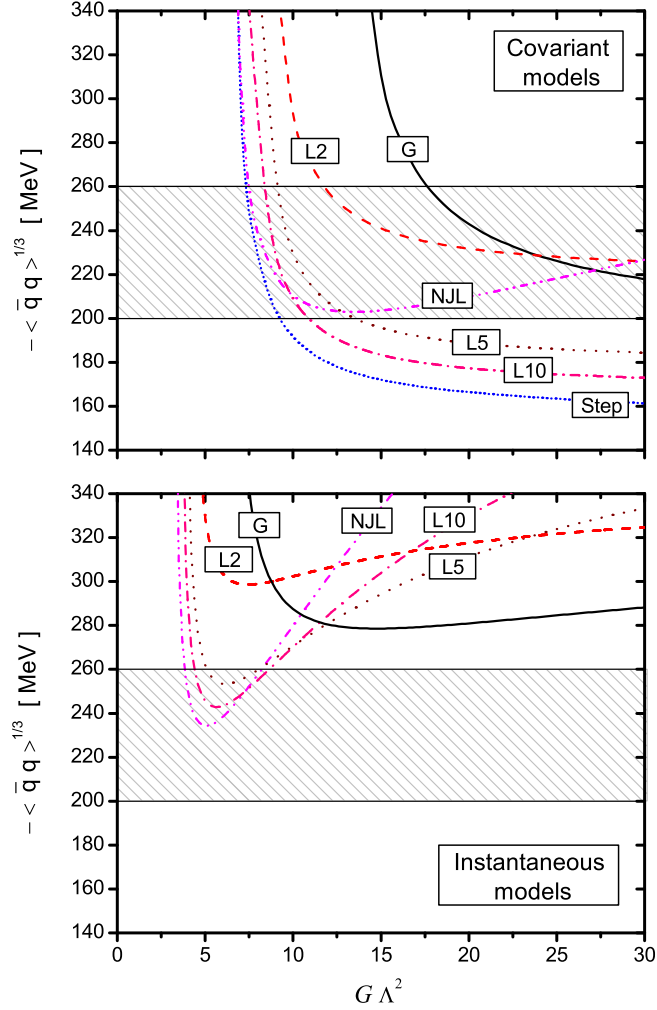} } \caption{Chiral
condensate as a function of $G \Lambda^2$ for different models in the chiral
limit. Upper and lower panels correspond to covariant and instantaneous
interactions, respectively.} \label{param}
\end{center}
\end{figure}

Let us start by analyzing the results in the upper panel of Fig.~2. We
observe that in all cases it is possible to choose a value of $G\Lambda^2$
so that the corresponding condensate falls within the empirical range. For
the case of smooth form factors (G and L2) we note, however, that
condensates in the lower half of the empirical range require quite large
values of $G\Lambda^2$. An important observation is that when the form
factors become sharper (Ln with increasing n) the corresponding curves
approach the values obtained for the Step form factor. This means that
although Eq.~(\ref{fpich}) includes a term with a derivative of the form
factor, which naively should be divergent for ${\rm n} \rightarrow
\infty$, this limit is well defined for $f_{\pi,0}$.

On the other hand, as it can be seen in the upper panel of Fig.~2,
in these models the Step limit does not match the curve obtained
for the 4dNJL model, i.e.\ the NJL model regularized by a
four-dimensional covariant cutoff. This fact can be understood as
follows. From Eqs.~(\ref{cond}) and (\ref{fpich}), the explicit
form of the function ${\cal F}(G\,\Lambda^2)$ for a covariant
model is given by
\begin{equation}
{\cal F}(G \,\Lambda^2) \ = \ x^{-2/3}\; I_1^{1/3}(x)\; I_2^{-1/2}(x) \ ,
\end{equation}
where $x = \bar \sigma/\Lambda$, and
\begin{eqnarray}
I_1(x) &=& \frac{N_c}{4 \pi^2} \ \int_0^\infty dz \
\frac{z\;s(z)}{z\, + \,x^2\, s^2(z)} \ \ ,
\\
I_2(x) &=& \frac{N_c}{8 \pi^2} \ \int_0^\infty dz \ \frac{\; 2\,z\;s^2(z)\, -
\, z^2\, s(z) \, s'(z)}{[\, z\, +\, x^2\, s^2(z)\, ]^2} \ \ .
\label{deriv}
\end{eqnarray}
Here we have defined  $s(z) = g(z \Lambda^2) = [\, r(z
\Lambda^2)\, ]^2$, whereas $s'(z) = ds(z)/dz$. The value of $x$ depends
implicitly on the dimensionless parameter $G\Lambda^2$ through the gap
equation
\begin{equation}
1 = \ 2 \; G\Lambda^2 \; I_3(x) \ ,
\end{equation}
where
\begin{equation}
I_3(x) \ = \ \frac{N_c}{4 \pi^2} \int_0^\infty dz \
\frac{z\;s^2(z)}{z + x^2 s^2(z)}\ .
\end{equation}
In general, the integrals $I_{i=1,2,3}$ cannot be performed analytically.
However, they can be evaluated for the case of a Step form factor by
considering the function
\begin{equation}
s(z) \ = \ \left\{
\begin{array}{ccrcl}
  1 & {\qquad\rm for\qquad} & z & < & 1 - \alpha/2 \\
  (1 + \alpha/2 - z)\,\alpha^{-1} & {\rm for} &
  1 - \alpha/2 & < & z \; < \; 1 + \alpha/2 \\
  0 & {\rm for} & z & > & 1 + \alpha/2
\end{array}%
\right.
\end{equation}
and taking the limit $\alpha \rightarrow 0$. By doing this, we
find that in the Step limit the integrals $I_1$ and $I_3$ are
equal to each other, and they are also coincident with their 4dNJL
model counterparts (the latter can be found e.g.\ in
Ref.~\cite{Klevansky:1992qe}). The actual results are
\begin{eqnarray}
I_1^{\rm Step}(x) = I_1^{\rm 4dNJL}(x) = I_3^{\rm Step}(x) = I_3^{\rm 4dNJL}(x) =
\frac{N_c}{4 \pi^2} \left[ 1 - x^2 \ln\left( 1 + \frac{1}{x^2}\right)
\right] \ .
\end{eqnarray}
On the other hand, for $I_2$ we get
\begin{equation}
I_2^{\rm Step}(x) \ = \ I_2^{\rm 4dNJL}(x)\; +\; \Delta I_2(x) \ ,
\end{equation}
where
\begin{eqnarray}
I_2^{\rm 4dNJL}(x) = \frac{N_c}{4 \pi^2} \left[ \ln\left( 1 +
\frac{1}{x^2}\right) - \frac{1}{1+x^2} \right] \ , \qquad
\qquad \Delta I_2(x) = \frac{N_c}{4 \pi^2} \ \frac{1}{\,4\,(1+x^2)} \ \ .
\end{eqnarray}
Due to the presence of the $\Delta I_2$ contribution, the Step limit of
the covariant nonlocal models does not match the 4dNJL result (as shown in
Fig.~2). It is not hard to check that $\Delta I_2$ comes from the
integral of the term with the derivative in Eq.~(\ref{deriv}), which is
nonzero only in the ``surface'' region $1 - \frac{\alpha}{2} < z < 1 +
\frac{\alpha}{2}$. The contribution of this term vanishes in the 4dNJL
according to the usual regularization procedure~\cite{Klevansky:1992qe}.

It is interesting to compare the results obtained within covariant models
with those provided by instantaneous nonlocal
models~\cite{Schmidt:1994di,Grigorian:2006qe}. The latter are shown in the
lower panel of Fig.~2. We observe that, in contrast with covariant models,
instantaneous nonlocal theories with soft form factors such as G and L2 are
not able to yield quark condensates that fall within the empirical range.
This can be achieved only for rather sharp form factors, like e.g.\ L5.
Another important difference is that for instantaneous nonlocal models the
curves obtained for Ln do converge to the corresponding NJL results (i.e.\
the results obtained within the NJL model with three-dimensional covariant
cutoff, 3dNJL). It is worth to mention, however, that for this type of models
$f_\pi$ is frame dependent, owing to the noncovariance. Here, to obtain the
results displayed in the lower panel of Fig.~2 we have used the pion rest
frame, which is the only frame where low energy theorems, such as the GT
relation, are fulfilled.

\subsection{Finite current quark mass case}

In the case of finite current quark masses, the model parameters $m_c$,
$G$ and $\Lambda$ are usually determined by fitting the pion mass and
decay constant to their empirical values $m_\pi=139$~MeV and $f_\pi =
92.4$~MeV, and fixing the chiral condensate at some phenomenologically
acceptable value. Since this last quantity is not well determined, we
consider here four representative values within the empirical range,
namely $(-\langle\bar q q \rangle)^{1/3}\, =\, 200$, 220, 240 and 260 MeV.
It should be noticed that, in the case of finite $m_c$, for schemes I and
II one gets different expressions for the integrals in
Eqs.~(\ref{vamopincha}) and (\ref{integrals}), which determine the values
of $m_\pi$ and $f_\pi$. Therefore, the fitted model parameters are also
expected to be different in both schemes. Our results are shown in Table
I, where we quote the parameter values for schemes I and II, considering
different form factors and quark condensates. Interestingly, the current
quark mass $m_c$ is basically equal for both schemes, while the remaining
two parameters ($G\Lambda^2$ and $\Lambda$) might be rather different,
especially in the case of soft form factors and low absolute values of the
quark condensate. Perhaps the most striking case corresponds to the L2
form factor: for Scheme II it is not even possible to find a parameter set
leading to a condensate of either $-\langle\bar q q \rangle^{1/3}\, =\,
200$ MeV or $220$ MeV. For the Gaussian form factor (i.e.\ the other
``soft'' function considered here), although we do find compatible
parameter sets, the obtained values of $\Lambda$ in Scheme II for the case
of low condensates are too low to be reliable for phenomenological
applications.

In Table II we quote the values obtained for $\bar\sigma$ and
$g^2_{\pi\gamma\gamma}$, together with the position of the first pole of
the quark propagator, for both covariant schemes I and II, and different
form factors and quark condensates. As discussed in the
literature~\cite{Sti87,Bowler:1994ir,Plant:1997jr}, the absence of purely
imaginary poles in the quark propagator can be understood as a realization
of quark confinement (we are working here in Euclidean space, thus a
purely imaginary pole corresponds to a real pole in Minkowski space). We
remark that in the nonlocal covariant models studied here, instead of one
single purely imaginary pole ---as in the case of the usual NJL model---
one has in general an arbitrary number of purely imaginary and/or complex
poles~\cite{Bowler:1994ir,Plant:1997jr,General:2000zx,Scarpettini:2003fj}
(we denote by ``complex pole'' a pole which has {\it both} non-vanishing
real and imaginary components). In fact, previous analysis show that for
most applications only the first pole, i.e.\ the closest one to the
imaginary axis, is phenomenologically
relevant~\cite{Bowler:1994ir,Plant:1997jr,General:2000zx,Scarpettini:2003fj}.
In the case of Scheme~I, as one can see from Table II, among the cases
considered the first pole is complex only for a condensate of 200~MeV and
form factors G and L10. Moreover, the case of L10 is not
phenomenologically viable in view of the low value of $\bar\sigma$. Thus,
we conclude that if one wants to avoid (following confinement arguments)
the presence of low lying purely imaginary poles, the only acceptable
situation for Scheme~I is that which corresponds to a soft (Gaussian) form
factor and a low value of $-\langle\bar q q \rangle^{1/3}$. If a low lying
purely imaginary pole is admitted, then Scheme I with a Gaussian form
factor and somewhat larger absolute values of the condensate can be
accepted. Sharper form factors would be excluded, leading in all cases to
too small values of $\bar\sigma$. In the case of Scheme II, one observes
that sharp form factors also tend to yield small values of $\bar\sigma$,
the sole exceptions being the cases L5 and L10 with $-\langle\bar q q
\rangle^{1/3}\, =\, 200$ MeV. In these cases, however, the situation is
also problematic, since the first poles, although complex, have a very
small imaginary part. This implies the existence of a very low $\bar q q$
pseudo-threshold, which would require additional prescriptions to preserve
meson stability already at energies of a few hundreds of
MeV~\cite{Scarpettini:2003fj}. Given the above discussed limitations of
Scheme~II for soft form factors and low absolute values of the condensate,
we conclude that the sets corresponding to G and L2 form factors and
values of $-\langle\bar q q \rangle^{1/3}$ between 240 and 260~MeV would
provide the more acceptable results. Notice that for these limiting values
of the condensate the lowest pole turns out to be complex (``confinement''
situation) and purely imaginary, respectively.

Concerning the values obtained for $g^2_{\pi\gamma\gamma}$, it is
seen that in all cases they are compatible with the experimental
range, $g_{\pi\gamma\gamma}^2=(7.5\pm 0.5)\times
10^{-8}$~MeV$^{-2}$. Now in Table III we quote the parameter
values and the results obtained for $\bar\sigma$ and
$g^2_{\pi\gamma\gamma}$ in the 4dNJL model. We recall that in this
case the propagator has a single pole, located at the constituent
mass $\Sigma=m_c+\bar\sigma$ on the imaginary axis. Notice that,
as commented in the Introduction, the model does not provide a
proper description of the $\pi^0\to\gamma\gamma$ decay width.
Phenomenologically acceptable values of $g^2_{\pi\gamma\gamma}$
can be obtained instead if the cutoff is (ad-hoc) relaxed to
infinity in the triangle loop integrals.

\begin{table}
\caption{Model parameters for various models leading to some
representative values of the chiral condensate ($m_c$ and $\Lambda$ given
in MeV)} \label{tab1}
\begin{center}
\begin{tabular}{ccccccccc}
\ $-\langle\, q \bar q\,\rangle^{1/3}$ \ & \ Form factor \ &
\multicolumn{3}{c}{Scheme I} & \ \ \ & \multicolumn{3}{c}{Scheme II}
\\ \cline{3-5} \cline{7-9}
                 & \rule{0cm}{.4cm} & \ $m_c$ \ &  $G \Lambda^2$   & $\Lambda$      &
&                                     \ $m_c$ \ &  $G \Lambda^2$   & $\Lambda$      \\
\hline
200                  &     G        &  9.7      &   18.82          &   651.9        &
&                                      9.8      &   71.11          &   459.7        \\
\cline{2-9}
                     &     L2       &  9.7      &   12.45          &   539.9        &
&                                      ---      &    ---           &    ---         \\
\cline{2-9}
                     &     L5       &  9.6      &    8.71          &   799.0        &
&                                      9.6      &   13.38          &   660.2        \\
\cline{2-9}
                     &     L10      &  9.8      &    7.57          &   885.0        &
&                                      9.7      &   10.68          &   716.5        \\
\hline
220                  &     G        &  7.4      &   16.98          &   772.0        &
&                                      7.4      &   29.06          &   604.0        \\
\cline{2-9}
                     &     L2       &  7.4      &   10.98          &   642.2        &
&                                      ---      &    ---           &    ---         \\
\cline{2-9}
                     &     L5       &  7.4      &    8.40          &   924.4        &
&                                      7.4      &   10.34          &   790.3        \\
\cline{2-9}
                     &     L10      &  7.5      &    7.47          &  1019.3        &
&                                      7.4      &    9.06          &   843.2        \\
\hline
240                  &     G        &  5.8      &   15.82          &   902.4        &
&                                      5.8      &   20.65          &   752.2        \\
\cline{2-9}
                     &     L2       &  5.8      &   10.14          &   751.8        &
&                                      5.8      &   16.06          &   586.8        \\
\cline{2-9}
                     &     L5       &  5.8      &    8.20          &  1059.2        &
&                                      5.8      &    9.27          &   925.7        \\
\cline{2-9}
                     &     L10      &  5.8      &    7.39          &  1163.0        &
&                                      5.8      &    8.36          &   978.6        \\
\hline
260                  &     G        &  4.6      &   15.08          &  1042.2        &
&                                      4.6      &   17.53          &   903.4        \\
\cline{2-9}
                     &     L2       &  4.6      &    9.61          &   868.0        &
&                                      4.6      &   11.77          &   736.1        \\
\cline{2-9}
                     &     L5       &  4.6      &    8.07          &  1202.8        &
&                                      4.6      &    8.73          &  1067.7        \\
\cline{2-9}
                     &     L10      &  4.6      &    7.34          &  1315.6        &
&                                      4.6      &    7.98          &  1122.5        \\
\hline
\end{tabular}
\end{center}

\end{table}

\begin{table}
\caption{Values for $\bar\sigma$ and $g_{\pi\gamma\gamma}^2$ and location
of the first pole of the quark propagator for various models
($g_{\pi\gamma\gamma}^2$ given in MeV$^{-2}\times 10^{-8}$, other
quantities given in MeV)} \label{tab2}
\begin{center}
\begin{tabular}{ccccccccc}
\ $-\langle\, q \bar q\,\rangle^{1/3}$ \ & \ Form factor \ &
\multicolumn{3}{c}{Scheme I} & \ \ \ & \multicolumn{3}{c}{Scheme II} \\
\cline{3-5} \cline{7-9}
      & \rule{0cm}{.4cm} & \ \ \ $\bar\sigma $ \ \ \ &
    First pole       & \ $g_{\pi\gamma\gamma}^2$ \ & \ \ \ \
                         & \ \ \ $\bar\sigma $ \ \ \ &
    First pole  & \ $g_{\pi\gamma\gamma}^2$ \ \\
\hline
200                  &     G   &  318  & 180\,+\,454\,i  & 7.43 &
&                                1356  & 499\,+\,241\,i\, & 7.16 \\
\cline{2-9}
                     &     L2  &  296  &  284\,i  & 7.37 &
&                                 ---  &  ---  &  --- \\
\cline{2-9}
                     &     L5  &  181  &  191\,i  & 7.57 &
&                                 366  &  640\,+\,169\,i\, & 7.37 \\
\cline{2-9}
                     &    L10  &  149  &  875\,+\,131\,i  & 7.39 &
&                                 302  &  711\,+\,98\,i\,  & 7.24 \\
\hline
220                  &     G   &  282  &  356\,i  & 7.50 &
&                                 620  &  421\,+\,372\,i\, & 7.45 \\
\cline{2-9}
                     &     L2  &  259  &  259\,i  & 7.49 &
&                                 ---  &  ---  &  --- \\
\cline{2-9}
                     &     L5  &  173  &  180\,i  & 7.74 &
&                                 276  &  757\,+\,217\,i\, & 7.45 \\
\cline{2-9}
                     &    L10  &  145  &  152\,i  & 7.64 &
&                                 248  &  835\,+\,120\,i\, & 7.46 \\
\hline
240                  &     G   &  255  &  288\,i  & 7.57 &
&                                 424  &  288\,+\,510\,i  & 7.49 \\
\cline{2-9}
                     &     L2  &  233  &  237\,i  & 7.55 &
&                                 475  &  485\,+\,304\,i\, & 7.48 \\
\cline{2-9}
                     &     L5  &  166  &  153\,i  & 7.87 &
&                                 238  &  243\,i  & 7.52 \\
\cline{2-9}
                     &    L10  &  142  &  147\,i  & 7.96 &
&                                 221  &  968\,+\,142\,i\, & 7.57 \\
\hline
260                  &     G   &  235  &  255\,i  & 7.61 &
&                                 339  &  430\,i  & 7.52 \\
\cline{2-9}
                     &     L2  &  216  &  219\,i  & 7.65 &
&                                 330  &  323\,i  & 7.50 \\
\cline{2-9}
                     &     L5  &  160  &  172\,i  & 7.94 &
&                                 215  &  220\,i  & 7.61 \\
\cline{2-9}
                     &    L10  &  139  &  143\,i  & 8.11 &
&                                 204  &  1110\,+\,165\,i\, & 7.66 \\
\hline
\end{tabular}
\end{center}

\end{table}

\begin{table}
\caption{Model parameters and values of $\bar\sigma$ and
$g_{\pi\gamma\gamma}^2$ for representative values of the chiral condensate
in the 4dNJL model ($g_{\pi\gamma\gamma}^2$ given in MeV$^{-2} \times
10^{-8}$, $m_c$, $\Lambda$ and $\bar\sigma$ given in MeV)} \label{tab3}
\begin{center}
\begin{tabular}{ccccccc}
\ $-\langle\, q \bar q\,\rangle^{1/3}$ \ &   \  \   & \ $m_c$ \ & $G \Lambda^2$ &
                               $\Lambda$ & $\bar\sigma$ & $g_{\pi\gamma\gamma}^2$ \\
\hline
200                  &     &  ---  &  ---   &   ---   &   ---  & --- \\
\hline
220                  &     &  7.7  &  9.13  &  807.8  &  297.9  &  4.40 \\
\hline
240                  &     &  6.0  &  7.93  &  947.8  &  244.1  &  5.79 \\
\hline
260                  &     &  \ 4.7 \ & \ 7.43 \ & \ 1094.6 \ & \ 218.2 \ & \ 6.50 \ \\
\hline
\end{tabular}
\end{center}

\end{table}

\section{Summary and conclusions}

We have presented a comparative description of chiral quark models which
include nonlocal covariant four fermion interactions. These approaches
---inspired in features of the underlying QCD theory--- represent natural
extensions of the usual NJL model and allow to overcome some of
its difficulties, in particular those related with the
regularization of ultraviolet divergences. We have concentrated
here in two alternative ways of introducing the nonlocality, which
we have called Scheme I and Scheme II. Using a common notation, we
have derived the main expressions to obtain the values of the pion
mass, the pion decay constant and some important mean field
quantities in terms of the model parameters $m_c$, $G$ and
$\Lambda$. In addition, for both schemes we have shown the
validity of various well-known relations in the chiral limit.

We have performed numerical calculations for gaussian and lorentzian form
factors, studying the behavior of model parameters, constituent masses and
propagator poles for given values of the chiral $\bar qq$ condensate
within a phenomelogically acceptable range. The parameters have been
fitted so as to reproduce the empirical values of the pion mass and decay
constant. In the chiral limit, the results for covariant nonlocal models
have been compared to those obtained within instantaneous nonlocal models.
We have found that, in contrast to covariant models, the latter can be
hardly accommodated to yield phenomenologically acceptable values of the
$\bar qq$ condensate. In the case of finite current quark masses, it has
been found that sharp form factors such as L5 and L10 are also
problematic, leading to either too low values of constituent quark masses
or to the appearance of a low $\bar qq$ pseudothreshold, which would need
additional prescriptions to preserve meson stability. On the other hand,
the results seem to favor quark interactions driven by smooth form factors
(in our analysis, G and L2). In the case of Scheme II, large values of
$-\langle\bar qq\rangle$ are found to be preferred, since lower
condensates lead to too low values of the ``cutoff'' $\Lambda$. For Scheme
I, instead, lower absolute values of the condensate seem to be favored in
order to avoid relatively small constituent quark masses. As it is shown
in Table II, in some of these cases one finds that the first pole of the
quark propagator has both a real and an imaginary part, which can be
understood as a manifestation of confinement. This pole structure is a
characteristic of the form factor under consideration, and can be extended
to more involved approaches such as e.g.\ the instanton liquid model.
Finally, we have considered the case of the NJL model regularized by a
four dimensional cutoff (4dNJL), analyzing its deviation from a nonlocal
model with a step-like form factor.

\acknowledgements

The authors are glad to thank D.~Aguilera, D.~Blaschke and H. Grigorian
for useful comments and discussions. This work has been supported in part
by CONICET and ANPCyT (Argentina), under grants PIP 02368, PIP 6084, PIP
6009, PICT00-03-08580 and PICT02-03-10718.

\end{document}